# The cellular uptake mechanism of SPIONs: an in-vitro study

Kiran Vishwasrao[1], Yasmin Khan[2] and S. Radha[1]

[1]University of Mumbai, Vidyanagari, Kalina, Mumbai-400098, India

[2]Sophia College for Women, Bhulabhai Desai Road, Mumbai-400026, India

Abstract

The Superparamagnetic Iron Oxide Nanoparticles (SPIONs) of sizes ranging from 10-50 nm are being used in a large number of biological studies because of their peculiar characteristics of inducing local hyperthermia, MR imaging, specific targeting and drug delivery. An in-vitro study of the cytotoxicity and an understanding of the specific pathway of cellular uptake will enable manipulation of conditions for optimal cellular uptake of SPIONs for targeted therapy. The objective of the present study was to identify the endocytotic pathway through which the SPIONs are taken up by C6 glioma cells. The cells were pre-incubated with different concentrations of pharmacological inhibitors and then exposed to SPIONs for a few hours. The endocytosed particles were localized and quantitatively estimated using Perl's or Prussian Blue reaction. There was significant reduction in the uptake of SPIONs when incubated with the inhibitor indicating the uptake of nanoparticles is being inhibited. This reduction in SPION uptake was found to be dependent on the concentration of the inhibitors and also the nature of the inhibitors. By a systematic study of choosing various inhibitors, the data can be narrowed down to the final pathway that may be involved in the SPION uptake. The present preliminary investigation is expected to provide insight of testing the possible mechanisms with complementary techniques.

# INTRODUCTION

Nanoparticles have shown great versatility in their use depending on their nature and size. Superparamagnetic iron oxide nanoparticles (SPIONs) in particular have been extensively used in magnetic imaging, in targeted drug delivery and in inducing local hyperthermia [1]. Studies have shown that SPIONs are physiologically well tolerated and further they are sometimes surface coated with ligands to target specific cells or with polymers such as dextran or albumin to prevent their rapid removal from circulation [2]. Our studies have shown that SPIONs are rapidly taken up by the C6 cell line, which is a transformed line of astrocytic lineage, *in vitro* and that they show no toxicity upto 72 hours in culture [3]. The astrocytes are the major glial cells of the nervous system and are required for the proper functioning of the neurons.

In order to render the use of nanoparticles in biological systems, it is essential to understand and optimize their uptake within cells. This study is aimed at elucidating the pathway through which SPIONs are internalized by C6 cells *in vitro*. Due to the small size of these particles, it is expected that they would be taken up by an endocytotic mechanism. The most common method of endocytosis is receptor-mediated or clathrin coated endocytosis, where material to be taken up binds to a receptor on the surface and then the membrane is pinched inside in the form of a vesicle which is coated by clathrin [4]. This study investigates the mechanism of endocytosis using pharmacological inhibitors to specifically inhibit a particular pathway and then delineate its effect on the uptake of SPIONs. The methods of inhibition used were exposure to hypertonic medium, using sucrose and NaCl, which is known to inhibit clathrin mediated endocytosis and also exposure to monodansyl cadeverine (MDC) which is a specific inhibitor of receptor-mediated endocytosis. Both these modes of inhibition reduced the uptake of SPIONs by C6 cells implicating these pathways in the uptake of the nanoparticles

## METHODOLOGY

The C6 glioma cell line (ATCC™ No: CCL-107) is routinely maintained in DMEM containing 10% Foetal Bovine Serum. These cells are seeded at a density of 5000 cells onto coverslips for visual analysis or in 96 well plates for the biochemical estimation. For visual detection the C6 cells were incubated with SPIONs for 24 hours while for quantitative analysis exposure was for 1 hour. Following the incubation, the cells were washed thoroughly using PBS to remove unbound particles. The cultures were then either stained or processed for quantitative estimation of the internalized SPIONs using the Perl's Prussian blue reaction [5]. Briefly the cells were treated with HCl for 1 hour and then with 5% $K_4Fe(CN)_6$ for 10 minutes. The colour intensity was estimated colorimetrically at 690nm.

In the inhibition assay, the cultures were treated with varying concentrations of the pharmacological inhibitors for 10 minutes after which SPIONs were added for 1 hour to allow for endocytosis. The duration of incubation was maintained at 1 hour as the inhibitors showed toxicity at higher durations of exposure.

The SPIONs used in this study were magnetite nanoparticles ranging in size less than 10 nm. These nanoparticles were synthesized by a chemical co-precipitation route, maintaining the pH in the range 9-14, as described by Kim et al [6]. The phase purity and size estimation was made by X-Ray diffraction and was ascertained by TEM measurements. Room temperature magnetization measurements were made on a vibrating sample magnetometer (VSM).

**RESULTS AND DISCUSSION**

The synthesized magnetite nanoparticles in the size range 7-9 nm was found to show a superparamagnetic behaviour at room temperature with a saturation magnetization of around 20 emu/g. The uncoated SPION samples as received from I.I.T. Madras was studied on a PC-based pulsed field hysteresis loop tracer [7] and found to show a saturation magnetization of 20 emu/g and coercivity of 110 Oe.

The uptake of SPIONs was visualized following staining of the cultures using the Perl's Prussian blue technique. Fig 1 shows the nanoparticles taken up by the C6 cells following 24 hours of incubation. The particles were found predominantly in the perinuclear region.

Estimation of the internalized particles demonstrated a dose dependent increase in the uptake. By 24 hours, most of the particles were already endocytosed and further periods of incubation did not significantly increase the uptake. Fig 2 is a graph showing the quantitative estimation of internalized SPIONs. The positive control is the same amount of SPIONs estimated in the absence of C6 cells.

To delineate the pathway for uptake of SPIONs, the C6 cells were pre-treated with hypertonic solutions (both sucrose and NaCl were used) that inhibit all clathrin-mediated endocytosis and with MDC, a specific inhibitor of receptor-mediated endocytosis. Uptake of SPIONs in the presence of the inhibitors was allowed only for 1 hour since at higher doses these chemicals were per se toxic to the C6 cells and caused them to round off and detach from the culture surface.

Fig 3 shows that all the treatments led to significantly reduced uptake of SPIONs implicating receptor-mediated clathrin coated pathway as the mode of uptake of these nanoparticles. Whereas treatment with both sucrose and NaCl, which were used to make the medium hypertonic showed a dose dependent inhibition, MDC treatment with all the doses showed the same amount of inhibition suggesting that MDC even at its lowest concentration is able to block the pathway. None of these inhibitors was able to completely block the uptake of nanoparticles suggesting that there might be other mechanisms by which the particles may be internalized. Complementary studies are expected to throw light on the possible mechanisms of cellular uptake.

**Conclusion**

This study demonstrates that uptake of SPIONs by C6 glioma cells is by endocytotic pathways as the process is dependent on the dose of nanoparticles and also since it is inhibited by pharmacological agents that specifically inhibit the various endocytotic pathways. Knowledge of the uptake mechanisms can by exploited to optimize the rate and amount of internalized particles depending on the required applications.

**Acknowledgement**

One of the authors (SR) gratefully acknowledges the Grant received from the Department of Science and Technology, Govt of India under the Women Scientist scheme implemented at Sophia College, Mumbai during 2005-08 while KV acknowledges staff of various departments of the University of Mumbai since 2017. The authors thank the Principal, Sophia College for support and the students Ms Venissa Machado, Ms Maruti Mishra and the laboratory staff for help in the experiments. The authors express their gratitude to Prof M.S.Ramachandra Rao and his team of I.I.T. Madras, Chennai for their kind gift of SPIONs, synthesized and characterized and made available for this study.

**References**

[1] Berry C. C. and Curtis A. S. G. (2003)  Functionalisation of magnetic nanoparticles for applications in biomedicine   J. Phys. D, Appl. Phys., **36:**  R198–R206

[2] Gupta A. K and Gupta M. (2005) .Synthesis and surface engineering of iron oxide nanoparticles for biomedical applications Biomaterials **26:** 3995–4021,

[3] Machado Venissa (2009) M.Sc thesis submitted to University of Mumbai

[4] Doherty G.J. and McMahon H.T. (2009) Mechanisms of Endocytosis Ann. Rev. Biochem. **78**: 857–902

[5] Cengelli, F., Maysinger, D., Tschudi-monnet, F., Montet, X., Corot, C., Petri-fink, A., Hofmann, H., et al. (2006). Interaction of Functionalized Superparamagnetic Iron Oxide Nanoparticles with Brain Structures, **318,** 108-116.

[6] D.M.Kim, Y. Zhang, W. Voit, K. V. Rao and M. Muhammed, J. Magn. Magn. Mater.**225** (2001) 30

[7] PC based pulsed field hysteresis loop tracer*,* S. D. Likhite, Prachi Likhite and S. Radha, Proc. DAE Solid State Physics Symposium, Manipal, Dec 2010.

**Figure Captions**

Fig 1. Qualitative localization of SPIONs using Perl's Prussian blue staining following 24 hours of incubation

Fig 2 Quantitative estimation of SPIONs in C6 glioma cells

Fig 3 Graph showing the effects of endocytic inhibitors on SPION uptake

Fig 1

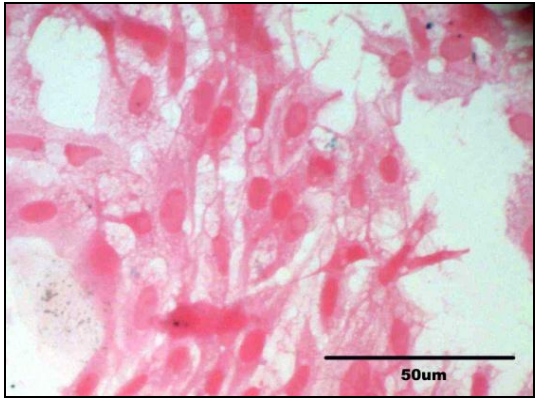 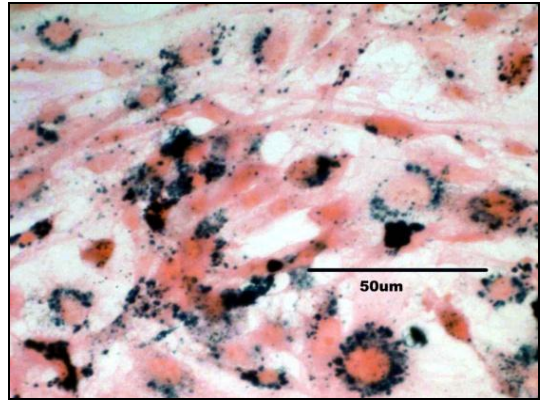

Control cells                                    Cells treated with 1000 µM SPIONs

Fig 2

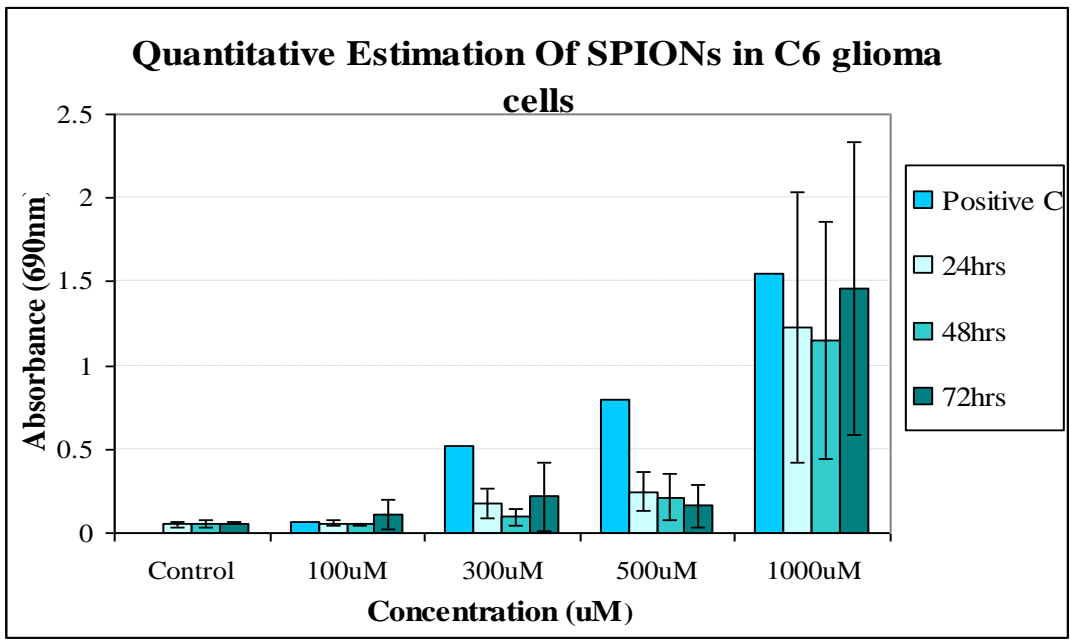

Fig 3

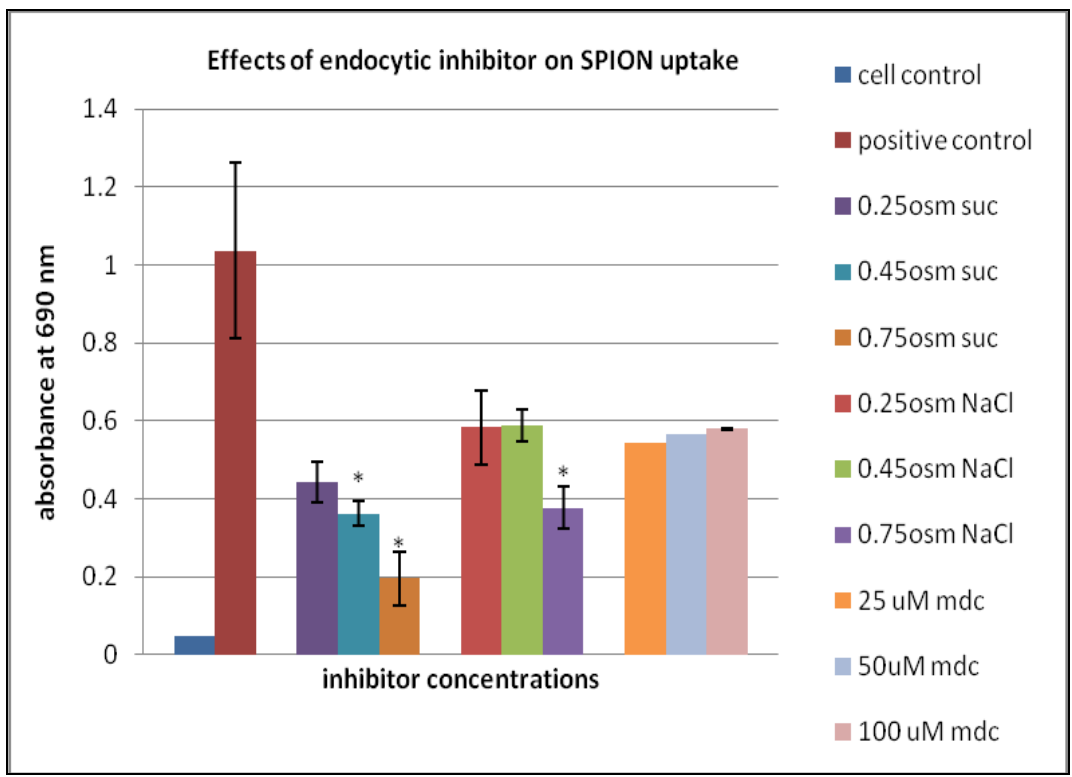